\documentclass{ws-p9-75x6-50}

\begin{document}

\title{Teleparallel Gravity and the Gravitational Energy-Momentum Density}

\author{V. C. de Andrade}

\address{D\'epartement d'Astrophysique Relativiste et de Cosmologie \\
Centre National de la Recherche Scientific (UMR 8629) \\
Observatoire de Paris, 92195 Meudon Cedex, France}

\author{L. C. T. Guillen and J. G. Pereira}

\address{Instituto de F\'{\i}sica Te\'orica,
Universidade Estadual Paulista\\
Rua Pamplona 145, 01405-900\, S\~ao Paulo, Brazil}

\maketitle

\abstracts{
In the context of the teleparallel equivalent of general relativity, we show that the
energy-momentum density for the gravitational field can be described by a true spacetime tensor.
It is also invariant under local (gauge) translations of the tangent space coordinates, but
transforms covariantly only under global Lorentz transformations. When the gauge gravitational
field equation is written in a purely spacetime form, it becomes the teleparallel equivalent of
Einstein's equation, and we recover M{\o}ller's expression for the canonical gravitational
energy-momentum pseudotensor.}

\section{Introduction}

One of the oldest and most controversial problems of gravitation is the definition of an
energy-momentum density for the gravitational field. As a true field, it would be natural to
expect that gravity should have its own local energy-momentum density. However, it is commonly
asserted that such a density can not be locally defined because of the equivalence
principle.\cite{gravitation} As a consequence, any attempt to identify an energy-momentum
density for the gravitational field leads to complexes that are not true tensors. The first
of such attempt was made by Einstein who proposed an expression for the energy-momentum
density of the gravitational field which was nothing but the canonical expression obtained
from Noether's theorem.\cite{trautman} Indeed, this quantity is a pseudotensor, an object
that depends on the coordinate system. Several other attempts have been made, leading to
different expressions for the energy-momentum pseudotensor for the gravitational
field.\cite{others}

It is usually accepted that in the context of general relativity, despite the existence of some
controversial points related to the formulation of the equivalence principle,\cite{synge} no
tensorial expression for the gravitational energy-momentum density can exist. However, as our
results show,\cite{prl} in the gauge context, the existence of an expression for the
gravitational energy-momentum density, which is a true spacetime and gauge tensor, turns out
to be possible. Accordingly, the absence of such expression should be attributed to the general
relativity description of gravitation, which seems to be not the appropriate framework to deal
with this problem\cite{maluf95}.

There has been along the years a continuous interest in this problem.\cite{recent} In particular,
a {\it quasilocal} approach has been proposed recently which is highly clarifying.\cite{nester}
According to this approach, for each gravitational energy-momentum pseudotensor, there is an
associated {\it superpotential} which is a hamiltonian boundary term. The energy-momentum
defined by such a pseudotensor does not really depend on the local value of the reference frame,
but only on the value of the reference frame on the boundary of a region --- then its {\it
quasilocal} character. As the relevant boundary conditions are physically acceptable, this
approach validates the pseudotensor approach to the gravitational energy-momentum problem. It
should be mentioned that these results were obtained in the context of the general relativity
description of gravitation.

In the present work, the gravitational energy-momentum problem is re-examined in the context of
teleparallel gravity.\cite{prl} Due to the fundamental character of the geometric structure
underlying gauge theories, the concept of currents, and in particular the concepts of energy and
momentum, are much more transparent when considered from the gauge point of view.\cite{gh}
Accordingly, we are going to consider gravity as described by a gauge theory.\cite{hehl} Our
basic interest will be concentrated on the gauge theories for the translation group,\cite{trans}
and in  particular on the so called teleparallel equivalent of general relativity.\cite{maluf}

\section{The Gravitational Energy-Momentum Density as a Gauge Current}

Let us consider the gauge gravitational field Lagrangian\cite{mgjg}
\be
{\cal L}_G = 
\frac{h c^{4}}{16 \pi G} \; S^{\rho \mu \nu} \; T_{\rho \mu \nu},
\label{gala}
\ee
where $h = {\rm det}(h^{a}{}_{\mu})$, $K_{\rho \mu \nu}$ is the contorsion tensor and
\[
S^{\rho \mu \nu} = - S^{\rho \nu \mu} \equiv {\textstyle \frac{1}{2}} \left[ K^{\mu \nu \rho}
- g^{\rho \nu} \; T^{\theta \mu}{}_{\theta} + g^{\rho \mu} \;
T^{\theta \nu}{}_{\theta} \right] 
\]
is a tensor written in terms of the Weitzenb\"ock connection only. As usual in gauge theories,
the lagrangian is quadratic in the field strength, represented here by the torsion tensor.
By using the relation
\be
\Gamma^{\rho}{}_{\mu \nu} = 
{\stackrel{\circ}{\Gamma}}{}^{\rho} {}_{\mu \nu} + 
K^{\rho}{}_{\mu \nu} ,
\label{relat}
\ee
the gauge lagrangian can be rewritten in terms of the Levi-Civita connection only. Up to a total
divergence, the result is the Hilbert-Einstein Lagrangian of general relativity
\be
{\cal L} = - \frac{c^4}{16 \pi G} \;  \sqrt{-g} \, {\stackrel{\circ}{R}} ,
\ee
where the identification $h = \sqrt{-g}$ has been made.

By performing variations in relation to the gauge field $A_a{}^\rho$, we obtain from
${\cal L}_G$ the teleparallel version of the gravitational field equation,
\be
\partial_\sigma(h S_a{}^{\sigma \rho}) - 
\frac{4 \pi G}{c^4} \, (h j_{a}{}^{\rho}) = 0 ,
\label{tfe1}
\ee
where $S_a{}^{\sigma \rho} \equiv h_{a}{}^{\lambda}
S_{\lambda}{}^{\sigma \rho}$. Analogously to the Yang-Mills theories,\cite{ramond} 
\be
h j_{a}{}^{\rho} \equiv \frac{\partial {\cal L}_G}{\partial h^a{}_{\rho}} = - 
\frac{c^{4}}{4 \pi G} \, h h_a{}^{\lambda} S_{\mu}{}^{\nu \rho} T^\mu{}_{\nu \lambda}
 + h_a{}^{\rho} {\cal L}_G
\label{ptem1} 
\ee
stands for the gravitational gauge current, which in this case represents the energy and
momentum of the gravitational field. The term $(h S_a{}^{\sigma \rho})$ is called
{\it superpotential} in the sense that its ordinary derivative yields the gauge current
$(h j_{a}{}^{\rho})$. Because of the anti-symmetry of $S_a{}^{\sigma \rho}$ in the last
two indices, $(h j_{a}{}^{\rho})$ is conserved as a consequence of the field equation:
\be
\partial_\rho (h j_a{}^\rho) = 0 .
\label{conser1}
\ee
Making use of the identity
\be
\partial_\rho h \equiv h {\Gamma}^{\nu}{}_{\nu \rho} =
h \left( {\Gamma}^{\nu}{}_{\rho \nu} - K^{\nu}{}_{\rho \nu} \right) \; ,
\label{id1}
\ee
this conservation law can be rewritten as
\be
D_\rho \, j_a{}^\rho \equiv \partial_\rho j_a{}^\rho +
\left( \Gamma^\rho{}_{\lambda \rho} - K^\rho{}_{\lambda \rho} \right) j_a{}^\lambda = 0 \; ,
\label{conser2}
\ee
where $D_\rho$ is the teleparallel version of the covariant derivative, which is nothing but
the Levi-Civita covariant derivative of general relativity rephrased in terms of the
Weitzenb\"ock connection.\cite{vector} As can be easily checked, $j_a{}^\rho$ transforms
covariantly under a general spacetime coordinate transformation, and is invariant under local
(gauge) translation of the tangent-space coordinates. This means that $j_a{}^\rho$ is a true
spacetime and gauge tensor. However, it transforms covariantly only under a {\it global}
tangent-space Lorentz transformation.  

\section{The Gravitational Energy-Momentum Pseudotensor}

To find the relation between the gauge approach and general relativity, it is necessary
to express the gauge field equation in a pure spacetime form. By considering the expression
of the Weitzenb\"ock connection in terms of the tetrad field,
\be
\Gamma^{\rho}{}_{\mu \nu}=h_a{}^\rho \partial_\nu h^a{}_\mu ,
\ee
Eq.(\ref{tfe1}) can be rewritten in the form
\be
\partial_\sigma(h S_\lambda{}^{\sigma \rho}) - 
\frac{4 \pi G}{c^4} \, (h t_{\lambda}{}^{\rho}) = 0 \; ,
\label{tfe2}
\ee
where now 
\be
h t_{\lambda}{}^{\rho} =
\frac{c^{4}}{4 \pi G} \, h \, \Gamma^{\mu}{}_{\nu \lambda} \, S_{\mu}{}^{\nu \rho}
+ \delta_\lambda{}^{\rho} \, {\cal L}_G
\label{ptem2} 
\ee
stands for the teleparallel version of the canonical energy-momentum pseudotensor of the
gravitational field. Despite not explicitly apparent, as a consequence of the {\it local}
Lorentz invariance\cite{weinberg} of the gauge Lagrangian ${\cal L}_G$, the field
equation~(\ref{tfe2}) is symmetric in $(\lambda \rho)$. Furthermore, by using the relation
(\ref{relat}), it can be rewritten in terms of the Levi-Civita connection only.
As expected, due to the equivalence between the corresponding lagrangians, it is the
same as Einstein's equation:
\be
\frac{h}{2} \left[{\stackrel{\circ}{R}}_{\mu \nu} -
\frac{1}{2} \, g_{\mu \nu}
{\stackrel{\circ}{R}} \right] = 0 \; .
\ee  

The canonical energy-momentum pseudotensor $t_{\lambda}{}^{\rho}$ is not simply the gauge
current $j_a{}^\rho$ with the algebraic index ``$a$'' changed to the spacetime index
``$\lambda$''. It incorporates also an extra term coming from the derivative term
of Eq.~(\ref{tfe1}):
\be
t_\lambda{}^\rho = h^a{}_\lambda \, j_a{}^\rho +
\frac{c^{4}}{4 \pi G} \, \Gamma^{\mu}{}_{\lambda \nu} S_{\mu}{}^{\nu \rho} \; .
\label{ptem3}
\ee
We see thus clearly the origin of the connection-term which transforms the gauge current
$j_a{}^\rho$ into the energy-momentum pseudotensor $t_\lambda{}^\rho$. Through the same
mechanism, it is possible to appropriately exchange further terms between the derivative and
the current terms of the field equation (\ref{tfe2}), giving rise to different definitions
for the energy-momentum pseudotensor, each one connected to a different {\it superpotential}
$(h S_\lambda{}^{\rho \sigma})$. Like the gauge current $(h j_a{}^\rho)$, the pseudotensor
$(h t_\lambda{}^\rho)$ is conserved as a consequence of the field equation:
\be
\partial_\rho (h t_\lambda{}^\rho) = 0 \; .
\label{conser3}
\ee
However, in contrast to what occurs with $j_a{}^\rho$, due to the pseudotensor character of
$t_\lambda{}^\rho$, this conservation law can not be rewritten with a covariant derivative.

Because of its simplicity and transparency, the teleparallel approach to gravitation seems
to be much more appropriate than general relativity to deal with the energy problem of the
gravitational field. In fact, M{\o}ller already noticed a long time ago that a
satisfactory solution to the problem of the energy distribution in a gravitational field
could be obtained in the framework of a tetrad theory. In our notation, his expression for
the gravitational energy-momentum density is\cite{moller}
\be
h t_\lambda{}^\rho = \frac{\partial {\cal L}}{\partial \partial_\rho h^a{}_\mu} \;
\partial_\lambda h^a{}_\mu + \delta_\lambda{}^\rho \, {\cal L} \; ,
\ee
which is nothing but the usual canonical energy-momentum density yielded by Noether's
theorem. Using for ${\cal L}$ the gauge Lagrangian (\ref{gala}), it is an easy task
to verify that M{\o}ller's expression coincides exactly with the teleparallel
energy-momentum density appearing in the field equation (\ref{tfe2}-\ref{ptem2}).
Since $j_a{}^\rho$ is a true spacetime tensor, whereas $t_\lambda{}^\rho$ is not,
we can say that the gauge current $j_a{}^\rho$ is an improved version of 
M{\o}ller's energy-momentum density
$t_\lambda{}^\rho$. Mathematically, they can be obtained from each other by
Eq.~(\ref{ptem3}). It should be remarked, however, that both of them transform
covariantly only under {\it global} tangent-space Lorentz transformations. This is,
we believe, the farthest one can go in the direction of a tensorial
definition for the energy and momentum of the gravitational field. The lack of a {\it local}
Lorentz covariance can be considered as the teleparallel manifestation of the pseudotensor
character of the gravitational energy-momentum density in general relativity. Accordingly,
we can say that, if it were possible to define a {\it local} Lorentz covariant gauge
current in the teleparallel gravity, the corresponding general relativity energy-momentum
density would be represented by a true spacetime tensor.

\section{Conclusions}

In the context of a gauge theory for the translation group we have obtained an
energy-momentum gauge current $j_a{}^\rho$ for the gravitational field which transforms
covariantly under spacetime general coordinate transformations, and is invariant under
local (gauge) translations of the tangent-space coordinates. This means essentially that
$j_a{}^\rho$ is a true spacetime and gauge tensor. By rewriting the gauge field equation in a
purely spacetime form, it becomes equivalent to Einstein's equation, and the gauge current
$j_a{}^\rho$ reduces to the canonical energy-momentum pseudotensor of the gravitational field,
which coincides with M{\o}ller's well-known expression. In the ordinary context of general
relativity, therefore, the energy-momentum density for the gravitational field will always be
represented by a pseudotensor.

By considering the {\it quasilocal} approach, we can say that to any energy-momentum
pseudotensor there is an associated {\it superpotential} which is a hamiltonian boundary
term.\cite{nester} On the other hand, the teleparallel field equations explicitly exhibit
both the {\it superpotential} and the gravitational energy-momentum complex. We see then
that, in fact, by appropriately exchanging terms between the {\it superpotential} 
and the current terms of the field equation (\ref{tfe2}), it is possible to obtain
different gravitational energy-momentum pseudotensors with their associated
{\it superpotentials}. In this context, our results can be rephrased according to the
following scheme. First, notice that the left-hand side of the field equation~(\ref{tfe2})
as a whole is a true tensor, though each one of its two terms is not. Then if we extract
the spurious part from the first term --- so that it becomes a true spacetime and gauge 
tensor --- and add this part to the second term --- the energy-momentum density --- it 
becomes also a true spacetime and gauge tensor. We thus arrive at the gauge-type field 
equation~(\ref{tfe1}), with $(h S_a{}^{\sigma \rho})$ as the {\it superpotential}, whose 
corresponding expression for the conserved energy-momentum density for the gravitational 
field, given by $j_a{}^{\rho}$, though transforming covariantly only under a global
tangent-space Lorentz transformation, is a true spacetime and gauge tensor. 

\section*{Acknowledgements}

The authors would like to thank FAPESP-Brazil, CAPES-Brazil and CNPq-Brazil for financial
support. V. C. de Andrade also thanks the D.A.R.C. - Observatoire de Paris, for the warm
hospitality.

\end{document}